\documentclass[fleqn,twoside]{article}
\usepackage[headings]{espcrc2}

\readRCS
$Id: espcrc2.tex,v 1.2 2004/02/24 11:22:11 spepping Exp $
\usepackage{graphicx}
\usepackage{psfrag}
\usepackage{amsmath}
\usepackage{amssymb}
\newcommand{\widebar}[1]{%
   \mkern1.5mu\overline{\mkern-1.5mu#1\mkern-1.mu}\mkern1.mu}
\newcommand{\eVdist}{\kern-0.06667em}
\newcommand{\gev}{{\,\text{Ge}\eVdist\text{V\/}}}
\newcommand{\fig}[1]{Fig.~\ref{fig-#1}}
\newcommand{\nb}{\,\text{nb}}
\title{Heavy flavour production in high-energy \(ep\) collisions}

\author{I.~Katkov\address{Skobeltsyn Institute of Nuclear Physics,
        Moscow State University,\\
        Vorob'evy Gory, Moscow 119992, Russia}\thanks{
        Supported by the Russian Federal Agency for Science and Innovations
        and DESY.
}
}
       
\begin{document}

\begin{abstract}
A selection of recent results on heavy quark production at the HERA
collider by the H1 and ZEUS collaborations are presented
with a focus on charmonium production in DIS, charm fragmentation 
and beauty production.
\vspace{1pc}
\end{abstract}

\maketitle

\section{Introduction}

In the heavy quark production processes there is at least one hard scale
and hence perturbative QCD calculations are expected to be reliable.
%Perturbative QCD caluclations imply the factorisation of the parton 
%distributions, hard scattering matrix elements and the description of the 
%hadronisation process. 
At HERA, where electrons or positrons of \(27.5\gev\)
are collided with protons of \(920\gev\) (\(820\gev\) before year 1998), 
pQCD predictions are tested in \(ep\) interactions.
The HERA collider is a unique facility to investigate 
the parton dynamics, the heavy quark production and fragmentation
properties and other topics.
%with important
%experimental input quantities: the parton distributions and
%fragmentation functions.

The main kinematic variables are the virtuality of the
exchanged photon, \(Q^2\), the photon--proton centre--of--mass energy,
\(W\), the Bjorken scaling variable, \(x\), and another scaling variable,
\(y\), which is the energy fraction of the lepton beam transferred to the
exchanged photon.

Both the photoproduction (\(Q^2<1\gev^2\)) and deep inelastic scattering
(DIS, \(Q^2\gtrsim 1\gev^2\)) regimes are studied.

\begin{figure}[htb]
%\vspace{9pt}
%\includegraphics[angle=90,width=20pc]{file}
\psfrag{aaa}%
{\bfseries\tiny\kern-1.1mm\({(}\!\text{C\kern-0.3mm S}\kern-0.3mm{+}%
\kern-0.3mm\text{C\kern-0.3mm O}\!{)}\)}
\includegraphics[width=75mm,height=58.6mm]{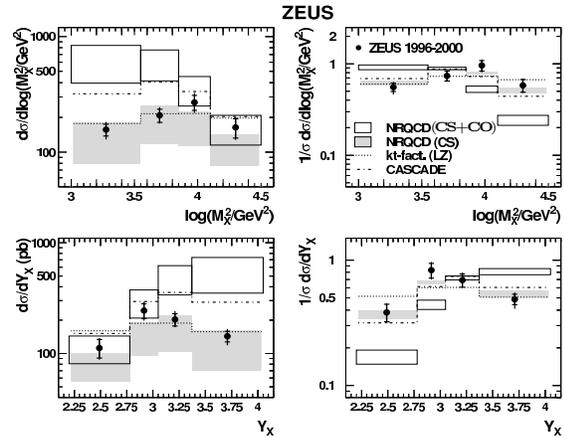}
%\framebox[55mm]{\rule[-21mm]{0mm}{43mm}}
\vspace*{-16mm}
\caption{
Absolute and normalised differential 
cross sections for the reaction \(ep\to eJ/\psi X\)
in the kinematic region 
\(2<Q^2<80\gev^2\), \(50<W<250\gev\),
\(0.2<z<0.9\) and 
\(-1.6<Y_\text{lab}<1.3\) as a function of
\( \log(M_X^2)\) %, (a), 
and \( Y_X \)%, (c)
.
%The inner error bars of the data points show the statistical uncertainty;
%the outer bars show statistical and systematic uncertainties added
%in quadrature.
The data are compared to LO NRQCD (\(\text{CS}{+}\text{CO}\)) predictions, 
a LO NRQCD (CS) calculation,
a prediction in the \(k_T\)--factorisation approach within the CSM %colour-singlet model
and the {\sc Cascade} (CCFM-based) MC predictions.
%Figures (b) and (d) show the data and the theoretical predictions
%normalised to unit area.
}
\label{fig-1}
\end{figure}

\section{Inelastic \(J/\psi\) production}

In \fig{1} recent results on the inelastic \(J/\psi\) production
in DIS regime  measured by ZEUS~\cite{ZEUS_jpsi_05}
%in the kinematic region \(2<Q^2<100\gev^2\), \(50<W<250\gev\), 
%\(0.2<z<0.9\)~\footnote{In the proton rest frame \(z\) is the
%exchanged photon energy fraction transferred to the \(J/\psi\).}
are compared to different theoretical predictions
including colour singlet (CS) and colour octet (CO) contributions as well as the CS
contributions only. Neither shapes nor normalisation are described if
CO contributions are included whereas CS contributions
alone generally agree with the data. This is also true for the 
properties of the hadronic final state (invariant mass, \(M_X\), and rapidity, \(Y_X\)) 
which were measured in this process
for the first time. From similar studies in photoproduction~\cite{ZEUS_jpsi_03} 
it is known that resummation techniques for soft gluon emission
can improve the theoretical description, however such calculations
are not available in DIS. \textsc{Cascade} Monte Carlo predictions 
are above the data but the shapes are well described.
A calculation in the \(k_T\)-factorisation approach
based on BFKL evolution equations gives the best description of the data.
\pagebreak

\section{Charm fragmentation}

\begin{figure}[htb]
%\vspace{9pt}
%includegraphics[angle=90,width=20pc]{file}
\includegraphics[width=75mm,height=60mm]{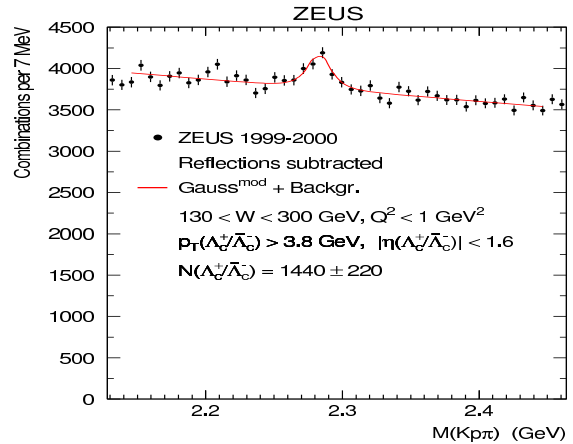}
%\framebox[55mm]{\rule[-21mm]{0mm}{43mm}}
\vspace*{-16mm}
\caption{
The \(M(Kp\pi)\) distribution
for the \(\Lambda_c^{+}\)/\(\widebar{\Lambda}_c^{-}\) candidates 
in photoproduction.
}
\label{fig-2}
\end{figure}

Fragmentation ratios and fractions in photoproduction
has been analysed by H1~\cite{H1_charm_05} and ZEUS~\cite{ZEUS_charm_05}. 
In this analysis the
charm ground states, \(D^0\), \(D^{+}\), \(D_s^{+}\), \(\Lambda_c^{+}\),
and the charm vector meson \(D^{*+}\)
were measured. %in the kinematic region \(Q^2<1\gev^2\), 
%\(130<W<300\gev\), \(p_T(D,\Lambda_c)>3.8\gev\), \(|\eta(D,\Lambda_c)|<1.6\).
As an example in \fig{2} the three-track invariant mass \(M(Kp\pi)\) distribution
is shown for the \(\Lambda_c^{+}\)/\(\widebar{\Lambda}_c^{-}\) candidates after
subtraction of reflections from \(D^{+}\) and \(D_s^{+}\) decays.
In charm fragmentation, the ratio of neutral to charged D-meson rates,
\(R_{u/d}=(cu)/(cd)=1.100\pm0.078(\text{stat.})\), is consistent with the
isospin invariance, the strangeness-suppression factor is
\(\gamma_s=(2cs)/(cd+cu)=0.257\pm0.024(\text{stat.})\) and 
the fraction of charged D mesons produced in a vector state
is \(P_V^d=0.566\pm0.025(\text{stat.})\)
which is not consistent with
the naive spin counting~\cite{ZEUS_charm_05}. 
The fragmentation fractions, \(f(c\to D,\Lambda_c)\),
were also measured~\cite{ZEUS_charm_05}: \(0.217\pm0.014\) for \(D^{+}\), \(0.523\pm0.021\) for
\(D^0\), \(0.095\pm0.008\) for \(D_s^{+}\), \(0.144\pm0.022\) for \(\Lambda_c^{+}\)
and \(0.200\pm0.009\) for \(D^{*+}\)~\footnote{Only statistical
errors are shown for the fragmentation ratios and fractions; all others
are omitted.}. These values are consistent with H1 results~\cite{H1_charm_05}.
Also they are in general agreement
and have precision competitive 
with the combined \(e^{+}e^{-}\) data~\cite{Gladilin_99} 
%and measurements previously 
%done by H1 in DIS regime~\cite{H1_charm_05} 
thus supporting the universality of
\(c\)-fragmentation.

A study of the charm fragmentation function have been done by H1
in DIS~\cite{H1_charm_05a} (\(2<Q^2<100\gev^2\), \(0.05<y<0.7\)). 
In the definition of the fractional transfer of the \(c\) quark energy to a \(D^{*}\) meson 
(\(D^{*}\to K\pi\pi_s\), \(p_T>1.5\gev\), \(|\eta|<1.5\))
the \(c\) quark energy was approximated by the energy in the appropriately 
defined hemisphere associated to the \(D^{*}\) in the \(\gamma^{*}p\)  frame:
\(z_{hem}=(E+p_L)_{D^{*}}/\sum_{hem}(E+p_L)\). This was done to establish
a close analogy to the measurements in \(e^{+}e^{-}\) experiments.
The distribution over \(z\) is in agreement with recent \(e^{+}e^{-}\)
annihilation data from CLEO at a similar centre-of-mass energy of the 
\(c\widebar{c}\) pair. The data are well described by \textsc{Rapgap}
Monte Carlo where either Peterson or Kartvelishvili parametrisation
of the fragmentation function was implemented. The best description
was achieved with 
\(\varepsilon=0.018_{-0.004}^{+0.004}\)
(Peterson)
or \(\alpha=5.9_{-0.6}^{+0.9}\) (Kartvelishvili).
The \(\varepsilon\) value is lower than
the previous ZEUS photoproduction result
\(\varepsilon=0.064\pm0.006_{-0.008}^{+0.011}\)~\cite{ZEUS_charm_02}.

\section{Charm with jets}

%\enlargethispage{\baselineskip}
Heavy quark production in association with jets 
measured recently by H1~\cite{H1_charmj_05} and 
ZEUS~\cite{ZEUS_charmj_05} provides a test of a multiscale process.
The ZEUS measurement of jet cross sections in charm 
photoproduction (\(Q^2<1\gev^2\),
\(130<W<280\gev\)) has been confronted with 
both massive and massless calculations.
%the massive
%calculation by Frixione et al. (FMNR)
%and the massless one by Heinrich and Kniehl.
In the so-called massive scheme there are three active quark
flavours in the proton or photon with heavy quarks
produced dynamically via  e.g.\ boson gluon fusion.
The approach is 
expected to be valid at low transverse momenta of the \(c\) quark. 
In the resummed massless scheme
there are up to five active partons and it is valid at high
transverse momenta. Combined calculation schemes 
(like variable flavour number scheme) also exist.

The \(D^{*}\) mesons (\(D^{*}\to K\pi\pi_s\))
were  required to be in the kinematic region \(p_T>3\gev\),
\(|\eta|<1.5\).
Jets (\(k_T\) cluster algorithm)
with \(-1.5<\eta^\text{jet}<2.4\), \(E_T^\text{jet}>6\gev\) were
selected. For the dijet analysis the highest \(E_T^\text{jet}\) jet
was required in addition to satisfy \(E_T^\text{jet}>7\gev\).

\begin{figure}[htb]
%\vspace{9pt}
%includegraphics[angle=90,width=20pc]{file}
\includegraphics[width=75mm,height=51.5mm]{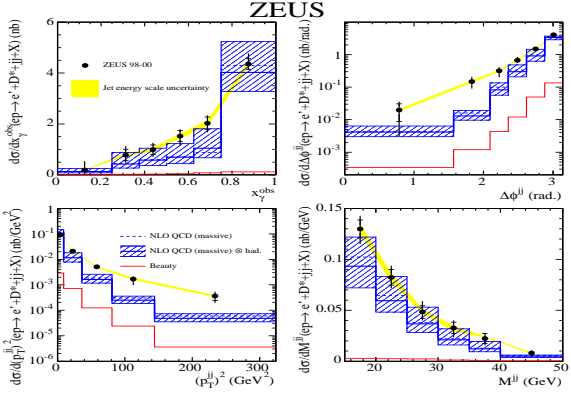}
%\framebox[55mm]{\rule[-21mm]{0mm}{43mm}}
\vspace*{-13mm}
\caption{
Dijet correlations in charm photoproduction 
compared to massive NLO QCD predictions.
The beauty component in the data (\textsc{Herwig} MC estimate) 
is also shown.
}
\label{fig-3}
\end{figure}

For all \(E_T^\text{jet}\) and \(\eta^\text{jet}\) the shapes of inclusive jet cross 
sections with and without a \(D^{*}\) tag
%(done in the \(\eta-\phi\) plane) 
are well described by both theoretical predictions
whereas the normalisation is described by their upper limits. There is 
no excess in the forward region (positive \(\eta^\text{jet}\)) observed
previously in the purely inclusive jet measurements. 
In \fig{3} dijet correlations are shown in comparison to
the massive NLO QCD predictions: the fraction of the photon momentum
participating in the dijet production, \(x_\gamma^\text{obs}\); 
the azimuthal separation of the jets, \(\Delta\phi^\text{jj}\);
the transverse momentum squared of the jets, \((p_T^\text{jj})^2\);
and the dijet invariant mass, \(M^\text{jj}\). The dijet correlations
are directly sensitive to the higher order corrections as
for the lowest order \(2\to2\) process \(\Delta\phi^\text{jj}=\pi\)
and \((p_T^\text{jj})^2=0\). The massive predictions are
low in comparison to the data at low \(\Delta\phi^\text{jj}\) and
high \((p_T^\text{jj})^2\). This discrepancy is enhanced for the
resolved-enriched (\(x_\gamma^\text{obs}<0.75\)) sample.
The \textsc{Herwig} MC describes the shape of the 
measurements well. This indicates  that either even higher-order
calculations or a matching of the NLO matrix elements to
a parton shower program (MC@NLO) are needed.

\begin{figure}[!h] %[hbt]
%\vspace{9pt}
%includegraphics[angle=90,width=20pc]{file}
\includegraphics[width=75mm,height=40mm]{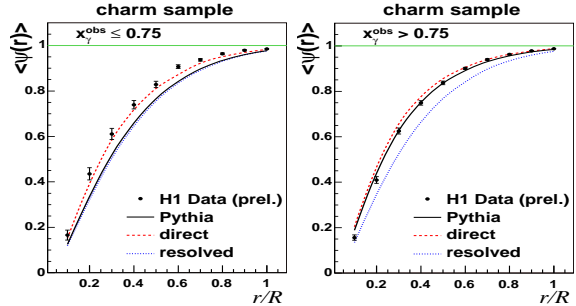}
%\includegraphics*[angle=270,width=75mm]{zzzplots/H1prelim-05-077.fig1a.eps}
%\framebox[55mm]{\rule[-21mm]{0mm}{43mm}}
\vspace*{-13mm}
\caption{
Average integrated jet shape for jets without charm tag for two different
regions of \(x_\gamma^\text{obs}\) (resolved- and direct-enriched). The data are compared to
\textsc{Pythia} MC predictions (total, direct and resolved contributions).
}
\label{fig-4}
\end{figure}

%\enlargethispage{\baselineskip}
Jet shape variables in charm photoproduction (\(Q^2<1\gev^2\), \(0.2<y<0.8\)) 
have been analysed
by H1~\cite{H1_charmj_05a}. An insight into the hard scatter process
is possible due to the fact that the integrated jet shape 
\(\psi(r)\) is expected to be different for quark- and gluon-initiated jets. 
Jets were reconstructed with the \(k_T\) cluster algorithm.
In the dijet sample (\(p_T^{{jet}_{1(2)}}>7(6)\gev\), \(|\eta|<1.7\)) 
one jet was tagged by a semileptonic decay muon 
(\(p_T^\mu>2.5\gev\))
as being initiated by a \(c\) quark using the two-dimensional fit of the muon
transverse momentum with respect to the closest jet, \(p_T^{rel}\), and the muon impact parameter,
\(\delta\), distributions to a mixture of ones in 
\textsc{Pythia} MC for light, charm and beauty quarks. 
The cut \(p_T^{rel}<1\gev\) was imposed to get a
charm enriched event sample. %(\(73\pm3\%\)). 
In \fig{4} the average integrated jet shapes \(\langle\psi(r)\rangle\) are shown for the
jet without the charm tag.  
For the resolved enriched sample (\(x_\gamma^\text{obs}\leqslant0.75\)) gluon
jets are expected to dominate via the resolved photon
charm excitation process, \(c_\gamma\,g\to c\,g\), according to the \textsc{Pythia} 
MC predictions. This dominance is not observed in the data and on the contrary
the integrated jet shape \(\langle\psi(r)\rangle\) is similar to one in the
direct enriched sample (\(x_\gamma^\text{obs}>0.75\)) where quark jets 
are expected to dominate via the photon gluon process,
\(\gamma\,g\to c\widebar{c}\).

\section{\(F_2^{c\widebar{c}}\) and \(F_2^{b\widebar{b}}\)}

%\enlargethispage{\baselineskip}
The inclusive lifetime tagging technique have been used by H1 
for the identification of heavy quark hadrons.
The method is based on the track impact parameter, \(\delta\),
as measured by the H1 vertex detector. For heavy quarks tracks
have high significance, \(S\), defined as \(\delta\) divided by its
error. Hence significance distributions \(S_i\) are considered
where \(S_i\) is the significance of track with \(i\)-th highest
absolute significance in an event. The \(c\), \(b\) and light quark 
fractions are extracted from the simultaneous fit to at least
two distributions \(S_1\) and \(S_2\). The
method features low extrapolation factors (high acceptances).
%as e.g{.} no high transverse momentum semileptonic decay
%muon is required to identify heavy quarks. 

\begin{figure}[htb]
%\vspace{9pt}
%includegraphics[angle=90,width=20pc]{file}
\includegraphics[width=75mm,height=81.2mm]{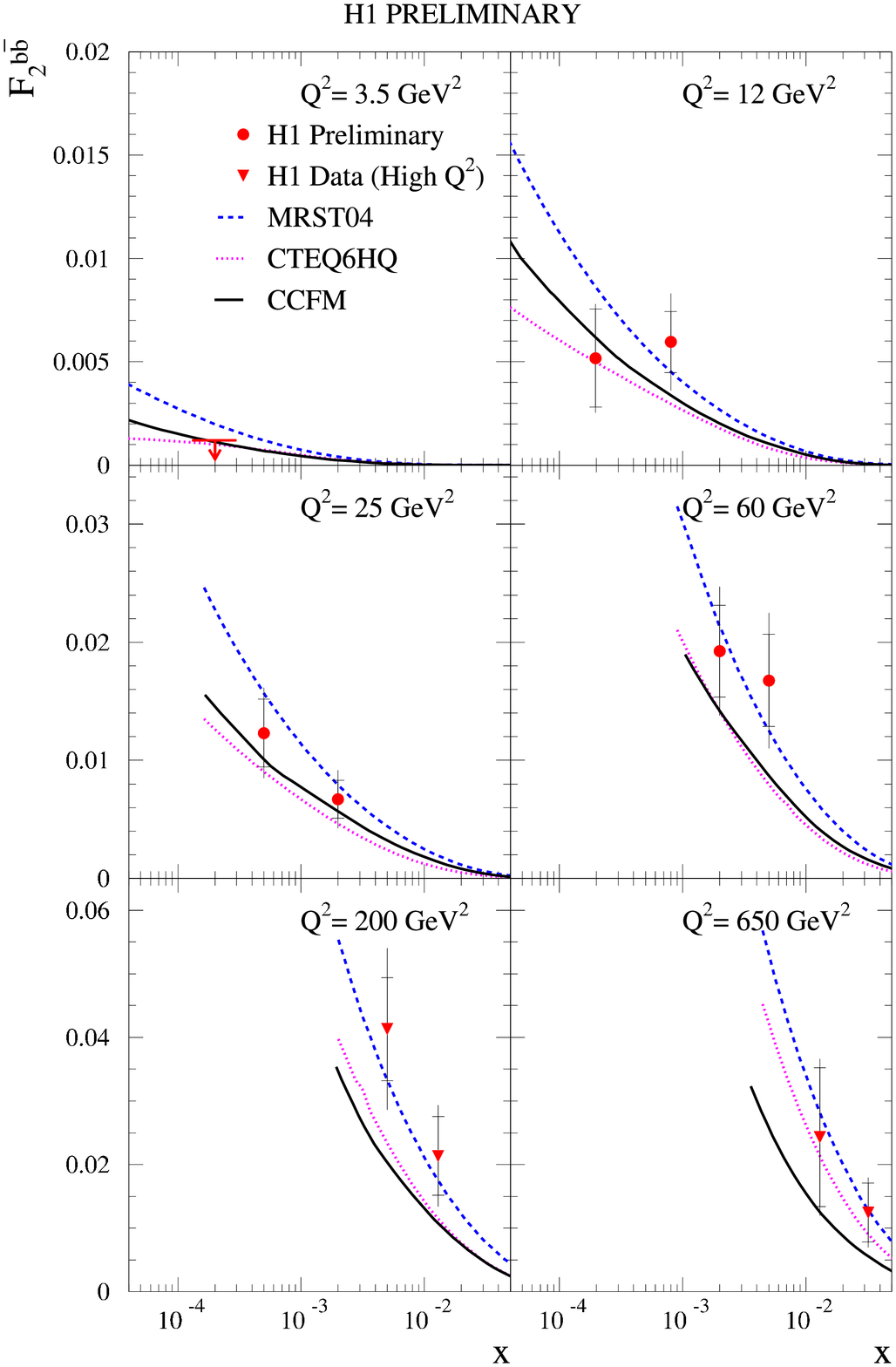} %height=80mm
%\includegraphics*[angle=270,width=75mm]{zzzplots/H1prelim-05-077.fig1a.eps}
%\framebox[55mm]{\rule[-21mm]{0mm}{43mm}}
\vspace*{-14mm}
\caption{
The structure function \(F_2^{b\widebar{b}}\) as a function of \(x\)
for different values of \(Q^2\). The data are compared to 
NLO predictions in the framework of the VFNS
(MRST, CTEQ) and predictions based on CCFM parton evolution.
}
\label{fig-5}
\end{figure}

The method allowed a  purely inclusive measurement of charm
and beauty dijet cross sections~\cite{H1_cbj_05} %\(\sigma(e p\to e Q \widebar{Q} X\to e j j X)\)
as well as the first measurement
of the beauty contribution in the structure function. %~\cite{H1_f2bb_05}.
Using three distributions \(S_1\), \(S_2\) and \(S_3\),
the measurement of \(F_2^{b\widebar{b}}\) shown
in \fig{5} was extended to \(Q^2\) as
low as \(Q^2\leqslant 60\gev^2\)~\cite{H1_f2bb_05a}. 
%There is no
%evidence for large excess of \(b\) quark production
%compared to 
The data are described by the
variable flavour number scheme (VFNS) NLO
calculations. In the kinematic region of the measurement
the beauty cross section is on average \(0.8\%\)
of the total \(e p\) cross section (\(2.7\%\) in the high-\(Q^2\)
measurement).

\section{Beauty with dimuons}

Specific topologies of events with \(B\) decays
allow the identification of heavy quarks.
Following the strategy of the previously done 
\(\mu{-}D^{*}\) correlation analyses,
the dimuon charge correlations and invariant
mass distributions have been studied by ZEUS to measure
beauty production~\cite{ZEUS_bmumu_05}. Despite rather restricted statistics
the analysis is sensitive to \(B\) mesons almost at rest
due to low background and high acceptance
down to low muon transverse momentum, \(p_T^\mu>1.5\gev\). 
The muon rapidity range, \(-2.2<\eta^\mu<2.5\), is
also wide due to almost full rapidity coverage of the muon
detectors used in the analysis. The azimuthal separation of the 
the muons, \(\Delta\phi^{\mu\mu}\), was measured 
for high dimuon invariant mass, \(m^{\mu\mu}>3.25\gev\), 
thus providing a direct probe of the \(b\widebar{b}\) correlations.
The total cross section measurement, 
\(\sigma_\text{tot}(ep\to b\widebar{b}X)=
16.1\pm1.8(\text{stat.})_{-4.8}^{+5.3}(\text{syst.})\nb\),
is to be compared to a NLO QCD prediction, \(6.8_{-1.7}^{+3.0}\nb\).

\section{Summary and outlook}

The large data sample accumulated during HERA I
running allowed H1 and ZEUS to test thoroughly
different aspects of QCD in charm and beauty
production processes. The NLO QCD calculations,
where available, were found to be in general agreement
with the data, although some measurements suggest
that even more precision is needed. 
New results from the 
post-upgrade HERA II running
with increased collision rates, polarised beams
and upgraded detectors will bring more experimental precision 
in the heavy flavour studies.
%are already under way.

\end{document}